# A Novel Scheme for Secured Data Transfer Over Computer Networks

**Rangarajan Athi Vasudevan**
(University of Michigan, Ann Arbor, USA
ranga@umich.edu)

**Ajith Abraham**
(School of Computer Science and Engineering, Chung-Ang University, Korea,
ajith.abraham@ieee.org)

**Sugata Sanyal**
(Tata Institute of Fundamental Research, India
sanyal@tifr.res.in)

**Abstract:** This paper presents a novel encryption-less algorithm to enhance security in transmission of data in networks. The algorithm uses an intuitively simple idea of a 'jigsaw puzzle' to break the transformed data into multiple parts where these parts form the pieces of the puzzle. Then these parts are packaged into packets and sent to the receiver. A secure and efficient mechanism is provided to convey the information that is necessary for obtaining the original data at the receiver-end from its parts in the packets, that is, for solving the 'jigsaw puzzle'. The algorithm is designed to provide information-theoretic (that is, unconditional) security by the use of a one-time pad like scheme so that no intermediate or unintended node can obtain the entire data. A parallelizable design has been adopted for the implementation. An authentication code is also used to ensure authenticity of every packet.

## 1 Introduction

Security of network communications is arguably the most important issue in the world today given the vast amount of valuable information that is passed around in various networks. Information pertaining to banks, credit cards, personal details, and government policies are transferred from place to place with the help of networking infrastructure. The high connectivity of the World Wide Web (WWW) has left the world 'open'. Such openness has resulted in various networks being subjected to multifarious attacks from vastly disparate sources, many of which are anonymous and yet to be discovered. This growth of the WWW coupled with progress in the fields of e-commerce and the like has made the security issue even more important.

A typical method for security that is used to prevent data from falling into wrong hands is encryption. Some encryption techniques like RSA [Rivest et al., 1978], which use asymmetric keys, involve algebraic multiplications with very large numbers. The cost that has to be paid in implementing encryption in networks is high owing to this computational complexity. While other techniques like the DES [FIPS, 46] which use symmetric keys are less secure computationally than their asymmetric counterparts.

Given the amount of computing power that is available, and considering also the growth of distributed computing, it is possible to break into the security offered by many such existing algorithms. So, any alternative to encryption is welcome so long as the level of security is the same or higher. Also, such an alternative should be more efficient in its usage of resources.

In practice, in a computer network, data is transferred across nodes in the form of packets of fixed size. Any form of security required is obtained by implementing cryptographic algorithms at the application level on the data as a whole. Then, the enciphered data is packetized at lower levels (in the OSI architecture) and sent. Any intruder able to obtain all the packets can then obtain the enciphered data by appropriately ordering the data content of each of these packets. Then, efforts can be made to break the cryptographic algorithm used by the sender. In the process of transmission, if it is possible to prevent any information release as to the structure of the data within the packets, an intruder would know neither the nature of the data being transferred nor the ordering of the content from different packets. This is what our algorithm achieves by using a one-time pad like scheme at the source.

The essential idea in our algorithm is to break the data that is to be transferred into many chunks, which are called parts [Vasudevan et al., 2004]. These parts when put together form the whole data but only if done so in a particular way, just like in a "jigsaw puzzle". The only way of doing so is known to the receiver for whom the data is intended. Any unauthorized node does not have enough information to carry out the right method of obtaining the parts from the packets and joining them, and then knowing it is correct (which is the property of the one-time pad). We have incorporated efficient techniques that realize the said security of the scheme, and at the same time provide scope for a fast, parallel implementation. The need for parallelization can only be understated owing to the demands of next generation processors and the growing popularity of dedicated hardware.

The rest of the paper is organized as follows. Some related research work is summarized in Section 2 followed by the technical details of the proposed algorithm in Section 3. Analysis, experiments and simulations are presented in Section 4. Some future research pointers and conclusions are provided in Sections 5 and 6 respectively.

## 2  Related Work

Owing to several issues mostly pertaining to key management, the theoretical one-time pad has been tough to implement practically. Numerous attempts have been made but under varying assumptions and conditions. One of the most recent has been [Xu et al., 2002] where one-time pads are used to protect credit card usage on the Internet. In [Maurer, 1999a], it has been argued that unconditional security can be obtained in practice using non-information –theoretically secure methods. This approach maintains that in the practical world, nobody can obtain complete information about a system owing to real-world parameters like noise. Likewise, [Cachin et al., 1997], [Maurer et al., 1999b] and [Günter et al., 2001] provide implementations of one-time pads and unconditional security but under assumptions about the environment and/or adversary. [Brassard et al., 1996] provides a quantum cryptographic view of one-time pads. Technological and practical limitations constrict the scalability of such methods. Chaotic maps are used to generate random numbers in [Fridrich, 1998], and these are used to build symmetric encryption schemes including one-time pads. Chaos theoretical methods though providing non-traditional methods of random number generation, are prone to cryptanalysis owing to the still-existent pseudorandom nature, and impose numerous restrictions on data size as is the case with [Fridrich, 1998].

Our approach provides an efficient implementation of the one-time pad without making assumptions or imposing restrictions, the likes of which are true of the references quoted in the previous paragraph. In the process, the core issues including key management are addressed and dealt with effectively. Due to its general nature, our algorithm can be deployed in most real-life networks without a fundamental change in the idea.

## 3 The Algorithm

We use the concept of Message Authentication Code (MAC) as suggested in [Rivest, 1998] to authenticate messages. For a packet of data, the MAC is calculated as a function of the data contents, the packet sequence number and a secret key known only to the sender and the receiver, and then it is appended to the packet. On receiving a packet, the receiver first computes the MAC using the appropriate parameters, and then performs a check with the MAC attached to the packet. If there is no match, then the receiver knows that the packet has been tampered with. A detailed explanation is provided in the next section.

Let the size of a packet in a network be denoted as *PS*. *PS* has a value of 1024-bits or 4096-bits typically. The prerequisite of the algorithm is the knowledge for the sender and the receiver only of a number *P*, exchanged a priori, of size $k*PS$, where *k* is a natural number. We can consider *P* in terms of blocks of size *PS* each, as $P_1, P_2, ..., P_k$. Thus *P* is the number obtained by the concatenation of the $P_i$ blocks for *i* from 1 to *k*, that is, *P* is $P_1P_2....P_k$.

In this algorithm, we only consider *k-1* parts of a data at a time, where each part is of any size less than or equal *PS - 2* (a detailed analysis of the algorithm is presented in the next section). If the entire data is not covered in these *k-1* parts, then the algorithm can be repeated by considering the next *k-1* parts and so on. Also, a random number of size *PS* is required to be generated for every *k-1* blocks processed. Denote this random number as *R*. The steps at the sender's end of the algorithm are as follows:

### S(D)

1) 'Tear the data D into N parts arbitrarily'. Consider the first $k-1$ parts of D. Call them $D_1, D_2, ... D_{k-1}$. Prefix and suffix each part by the binary digit '1'.

2) Perform the operation $1D_i1$ *XOR* $P_i$ from 1 to $k-1$. Denote them as $D_1'$, $D_2'$, ... , $D_{k-1}'$

3) Form $D_k'$ as $D_k' = R$ *XOR* $P_k$.

4) Perform *transform (P,R)*.

5) Generate a new random number and assign it to $R$.

6) Repeat steps 1, 2, 3, 4, 5 for the next $k-1$ blocks of data, and so on until all N parts are processed.

Now the packets actually transferred are formed from the $D_k'$ blocks, packet sequence number and the MAC (calculated as described earlier). At the receiver's end, the steps are as follows:

### R(D')

1) Perform a check on the MAC for each packet. If satisfied, GOTO next step.

2) Order packets according to the packet sequence number.

3) For each group of $k$ packets perform the following:

   a. Perform $P_i$ XOR $D_i'$ for all $i$ from 1 to $k-1$.

   b. Perform $P_k$ XOR $D_k'$ and obtain R.

   c. Remove the leading and trailing '1' of all the values obtained from the previous two steps.

   d. Perform *transform(P,R)*.

   *Continue*

The algorithm for the operation *transform(P,R)* is as follows:

*transform(P,R)*

1) Set $P_i \leftarrow P_i$ *XOR* $R$ for all $i$ from 1 to $k-1$.

2) Set $P_k \leftarrow P_k * R$.

For the transfer of the next data, the following is done. The sender knows the number of parts of the previous data and hence knows the value of $N \% k$, where % denotes the 'modulo' operation. Now, this value subtracted from $k$ provides the $P_i$'s unused in the last run of the loop in the algorithm. The next data to be transferred is first broken into parts as before. For the first run of the loop in the algorithm, the first $k - 1 - (N \% k)$ (here, this is the old value of N) parts are only considered and the run executed. For the remaining runs, we consider $k-1$ parts as before, and the algorithm is continued.

This is done for all subsequent data transfers between the sender and the same receiver.

## 4 An Analysis

### 4.1 Typical Encryption Algorithms

In this section, we briefly mention the algorithms of RSA and DES, an example each from the asymmetric and symmetric key encryption world respectively to later help understand the advantages of our algorithm over these typical encryption methods. The AES, which is also a symmetric key encryption algorithm but more secure than the DES, has not been discussed here but instead has been compared to our algorithm in a later sub-section to highlight the computational edge of our algorithm.

#### 4.1.1 DES

The DES has a 64-bit block size and uses a 56-bit key during execution (8 parity bits are stripped off from the full 64-bit key). It is a symmetric cryptosystem and was originally designed for implementation in hardware. When used for communication, both sender and receiver must know the same secret key, which can be used to encrypt and decrypt the message, or to generate and verify a message authentication code (MAC).
Though there have been two classes of attacks on the DES – the linear cryptanalysis in [Matsui, 1994], and the differential cryptanalysis in [Biham et al., 1993], both have proven to be impractical. Nevertheless, 56 bits keys are considered vulnerable to exhaustive search, which on an average would take $2^{55}$ steps. Therefore, NIST has recommended the use of AES in place of DES to ensure higher security.

#### 4.1.2 RSA

The RSA cryptosystem is a public-key cryptosystem that offers both encryption and authentication. It works as follows:
1) Take two large primes, $p$ and $q$, and compute their product $n = p * q$; $n$ is called the modulus.
2) Choose a number, $e$, less than $n$ and relatively prime to $(p - 1) * (q - 1)$.
3) Find another number $d$ such that $(e * d - 1)$ is divisible by $(p - 1) * (q - 1)$. The values $e$ and $d$ are called the public and private exponents, respectively. The public key is the pair $(n, e)$; the private key is $(n, d)$.
4) Take the message $m$ to be encrypted and calculate the ciphertext $c$ as $c = m^e \bmod n$, where $(n, e)$ is the public pair belonging to the receiver.
5) To decrypt, the receiver exponentiates the ciphertext to retrieve the message; that is, $m = c^d \bmod n$; the relationship between $d$ and $e$ ensures that the receiver gets the correct message.

RSA relies heavily on the operation of modular exponentiation, which is performed by a series of modular multiplications. It is due to this that RSA implementations are slower by a great margin than block ciphers like DES which is generally at least 100 times faster in software and between 1,000 and 10,000 times as fast in hardware depending on the implementation [RSALab, 2000].

The security of RSA relies on the computational difficulty of the prime factorization problem. To elucidate, given a number n, the problem is to find its prime factors which when multiplied with each other provide the same number. For a large n that obeys certain properties, this is almost impossible. However, in 1999, a 512-bit key (key refers to the modulus n) was factored after seven months of sustained effort [RSALab, 2000]. Coupled with the recent proof in [Agrawal et al., 2002] that the prime factorization problem can be solved in polynomial time complexity itself, the security of the RSA algorithm is likely to be tested in the future.

### 4.2 One-Time Pad

The one-time pad was first designed by Vernam in 1926 [Vernam, 1926], and its security was subsequently proven by Shannon in 1949 [Shannon, 1949]. It is constructed using a key chosen randomly which is at least as big as the message to be protected. Then, the key and the message are bitwise XORed to produce the ciphertext, which is sent to the receiver. The receiver, in possession of the key used by the sender, is able to bitwise XOR the ciphertext with this key and obtains the plaintext. Any intruder on intercepting the ciphertext can only guess the plaintext since the key is chosen randomly.

The one-time pad provides perfect secrecy and this is viewed as the benchmark to be attained by all cryptosystems. However, since a key once used cannot be used again owing to information leak, the one-time pad is practical only at great costs of key management. In this paper, we have provided an efficient implementation of the one-time pad including aspects of key management. The important difference between our implementation and that of the theoretical one-time pad is in the security offered under particular conditions. This has been discussed in detail in a later section.

### 4.3 A Discussion

The essential idea in the algorithm is to split the data into pieces of arbitrary size (rather, the size is not arbitrary since it is bound by 0 and *PS-2* but is allowed to take any value between the limits), transmit the pieces securely, and provide a mechanism to unite the pieces at the receiver's end. Towards this, the role of the number *P* is to provide a structure for the creation of the "jigsaw puzzle". This structure masks the pieces as well as protects the data within. The structure is changed periodically with the knowledge of both the sender and the receiver to prevent an adversary gaining information about it.

The XOR function is reversible and can be easily performed by the receiver since he knows the secret number *P*. However, an adversary without knowing *P* cannot obtain any additional information. This is because of the following.

Given a ciphertext *C* of length *PS* and a random secret key *P* of length *PS*, the probability of any particular key is the same. If it is possible to guess the message *M* such that *C = M XOR P*, then it is possible to determine the value of *P*. Since every secret key *P* is equally likely, there is no way of guessing which of the possible messages of length *PS* or less was sent. In other words, let us assume that the adversary has as much knowledge of the cryptographic mechanism as the receiver does (except, of course, knowledge of the secret key *P*). Now, the only knowledge that an adversary has about both *M* and *P* is that one is a binary string of length *PS* while the other is of length less than *PS*. Then, knowledge of the ciphertext *C* does not provide the adversary with any more information pertaining to either *M* or *P*. This is the information-theoretic security property of a one-time pad (for more details, refer [Shannon, 1949]). The value $D_i'$ obeys this one-time pad property since both $P_i$ and $D_i$ are random numbers as far as the adversary is concerned. Thus, it is impossible for an adversary to get any more information given this value. Thus, all of $D_i$ remain completely unknown to the adversary.

Now, after *k* parts are processed, it is not possible to repeat the values $P_1$, $P_2$, ...., $P_k$ Else, by manipulations due to the reversible nature of XOR, some information might be leaked to the adversary. Therefore, the next set of values has to be changed, and towards this a random number of equal size is used. Also, this random number is to be conveyed to the receiver without any adversary knowing its value. Hence, we introduce the random number as the *k*th part. This random number is used to calculate the next *P* to be used. Since the initial *P* was a secret for the adversary, and so is the random number, the new value of *P* is also a secret. Thus, the security of the data transmission is ensured. In our model of the one-time pad, data is of effective size less than *(k-1) \* PS* bits and the key XORed with the data at each stage is the number $P_1$, $P_2$,...$P_{k-1}$. The random number *R* is used as an input to a function (namely *transform()*) to generate the key for the next run. The value $P_k$ is used to securely convey the random number *R* that is generated by the sender, to the receiver.

The need for prefixing and suffixing every part with the bit '1' is explained below. By XORing a part with a random (with respect to the adversary) key of size *PS*, we are effectively embedding the data in some place in the key. However, there is a problem of the receiver not knowing where the data is embedded. This is best illustrated with a hypothetical example. Suppose the part is the binary sequence *01101*, and the key is the sequence *11000110* where PS = 8. Now embedding the data at position 3 from the left in the key, we get the value as *11011000*. When the receiver performs the XOR of this value with the key, the value he obtains is *00011010*. Now, there is no way of knowing whether the leading or the trailing zeros are part of the data or not. This is the reason for the use of the single bit '1' to prefix and suffix each part. As an illustration of this, we continue with the same example. Note that the part is necessarily smaller than the key by at least 2. Here the part under question is affixed and prefixed by '1', and then embedded in the key at position 2, say (position 3 can no longer be used). In this case, the resultant value is *10011101*. On performing the XOR with the key, the receiver obtains the value *01011011*. From this value, he considers only the binary sequence sandwiched between the first and last occurrences of the bit '1' which is the intended part.

A point of worry is that the arbitrary 'tearing' of data might result in a huge expansion of data. That is, due to the randomness in the splitting, the amount of data sent to convey some fixed amount of information might be huge. To avoid this, the algorithm provides the flexibility to fix a lower limit on the size a part can take. This ensures that a minimum amount of information is transferred to the receiver by each part. Thus, the overheads in the algorithm can be suitably controlled by the lower-limit size. It has been proved by in [Shannon, 1949] that in a one-time pad, as long as the data is of a size lesser than or equal to the key used, then the perfect-secrecy property is maintained. This justifies the statements made in this paragraph. In fact, for least overhead, each part could be made of size *PS* (in this case, there is no need to prefix and suffix with bit '1'). Our algorithm is however a general case of this, providing more security to lessen the probability of a serendipitically successful attack.

Our algorithm is a symmetric-key based security paradigm and hence it could be used for single-user security purposes as well, like for example file storage on disks. Note that unlike typical single-user encryption algorithms like DES, since state apart from key information is also required for the functioning of the algorithm, it should also be securely stored to, say, access the "jigsaw"-protected files subsequently.

### 4.4 Message Authentication

Message Authentication Codes (MACs) are tags attached to messages that are computed by an authentication scheme together with a secret key such that they can be verified by anybody possessing the same secret key and knowing the scheme. There are four broad classifications for MACs. We shall however mention only the two that interest us – unconditionally secure MAC, and hash function-based MAC.

A MAC is said to be *(q,e)-secure* if an adversary cannot construct a valid MAC for any new message with probability greater than *e*, given that the adversary has previously seen valid MACs for at most *q* messages. Unconditionally secure MACs of this type exist for any desired values of *q* and *e*. But, the lesser the parametric values are, the more stringent are the requirements on the inputs to the algorithm. For example, in order to have a MAC that spans a large data size (that is, large value for *q*), the key required should also be large.

So, in a practical setting, it is advisable to settle for cryptographic security rather than unconditional security. The modeling of these cryptographically secure MAC algorithms takes time as a parameter. That is, a MAC is said to be *(q,e,T)-secure* if an adversary cannot construct a valid MAC for any new message with probability greater than *e*, given that the adversary has previously seen valid MACs for at most *q* messages, and that his computation time is at most *T*. A particular class of MACs that we consider is the MACs based on hash functions like the HMAC ([Bellare et al., 1996]). In this class, owing to the nature of the design, the security offered by the MAC is normally the same as the security offered by the underlying hash function. Hence, these hash functions should be chosen with care to possess cryptographic properties such as resistance to collision detection and message authentication when applied to single blocks.

HMAC as defined in [Bellare et al., 1996] consists of the following step

$$H( K \text{ XOR } opad, H( K \text{ XOR } ipad, msg) )$$

where ipad, opad are constants and msg is the message to be authenticated. As can be seen, the essential computational effort as well as the security of the algorithm is primarily dependent on the hash function H. A good example of a strong cryptographic hash function is the SHA-1 [FIPS, 180-1]. An important point to note is that since MAC algorithms provide time-dependent security, it is important to frequently exchange keys between the same sender-receiver pair.

For the use of MACs in our algorithm, another secret key is required to be exchanged between the sender and the receiver as is required. However, since any *PS*-size block of *P* cannot be guessed from its first use, it is possible to use any of them or parts of it (in case a smaller key is desired). This information is also to be exchanged a priori along with the number *P* as well as the value of *k*. It is prudent to see that the same sender has different key arrangements with different receivers.

**4.5 Implementation Issues**

In our algorithm, for the transfer of N parts of data, there are N additions performed. Apart from these operations, the value of *P* is changed *floor(N/k)* times. At each change, there are *k-1* additions and 1 multiplication. Hence, in total our algorithm requires *N + floor(N/k) * (k-1)* additions and *floor(N/k)* multiplications on *PS*-bit blocks for the transfer of N parts of data. There is a degree of uncertainty involved in the amount of information transferred in a part as the splits happen arbitrarily. This uncertainty can, however, be removed marginally by imposing a lower limit on the size a part can take thus ensuring a transfer of minimum information in each part. Let the value of *k*=7. To transfer data of size 10 * *PS*, assuming the best case of *PS*-size parts, we have N = 10, and therefore we effectively require 16 * *PS* bit additions and 1 *PS*-bit multiplication. In the worst case, assuming the lower limit on size of each part to be *PS* / 2, we require 20 parts to transfer the same amount of data in which case we would effectively require 32 * *PS* bit additions and 2 *PS*-bit multiplications. A graph of the best case (corresponding to each part being of maximum size = *PS*) and the worst case (corresponding to each part being of minimum size = *PS* / 2) in the number of additions versus data size is shown in Figure 1.

A comparison with other encryption algorithms is valid only when the application of those algorithms is for secure data transmission. From this viewpoint, the statistics presented above compare favorably with respect to encryption algorithms like Advanced Encryption Standard (AES) [FIPS, 197]. In AES, for input block size of 128-bits and key length of 128-bits, there are at least 11 XOR operations apart from matrix multiplications, table lookups and vector shifts. When AES is scaled to an input of 10 *PS*-size blocks, it requires 110 * *PS* bit additions, far more operations than our algorithm functioning in the worst case as evident in Figure 1 (we have not included the matrix multiplications, table lookups and vector shifts in the calculations). Also, our algorithm does not transform the data except for the XOR operation. This operation and its inverse can be easily computed. Hence, as compared to encryption algorithms like AES, DES [FIPS, 46] and RSA [Rivest et al., 1978], the data processing time is least for our algorithm (the comparison with AES is graphically depicted in Figure 1).

Now, the algorithm has been designed to take into consideration *all* communication between particular sender-receiver pairs. Hence, it is practical to assume that the volume of the data typically exchanged between a sender-receiver pair is not small. Figure 2 depicts the influence of the size of the data (intentionally taken large) on the number of XORs for different values of *k*. Given the expression for the number of XORs as a function of *k* and data size, we see that the line corresponding to value of *k*=2 corresponds to the least number of XOR operations. In fact, for larger and larger data sizes, due to the smaller slope of this line as compared to other values of *k*, the value *k* = 2 is the optimum by a large margin. Figure 3 illustrates the variation in number of XORs over a wide range of values of *k* for different data sizes.

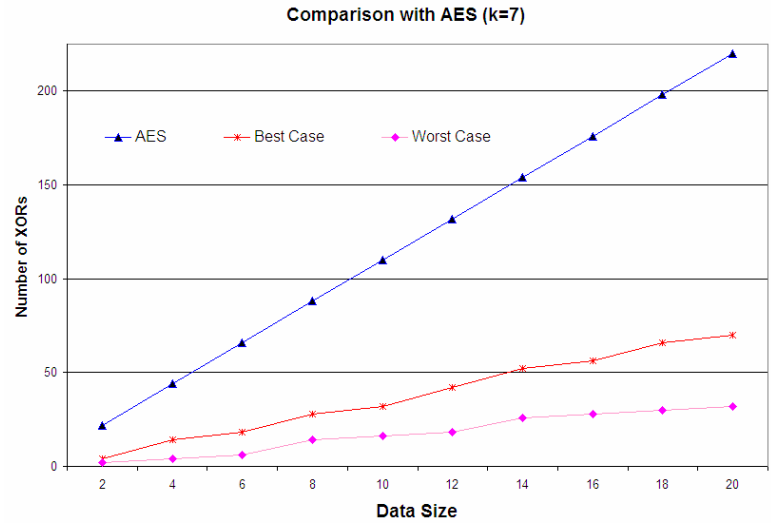

*Figure 1: Comparison of AES and our algorithm functioning in a) best case and b) worst case in the number of XORs*

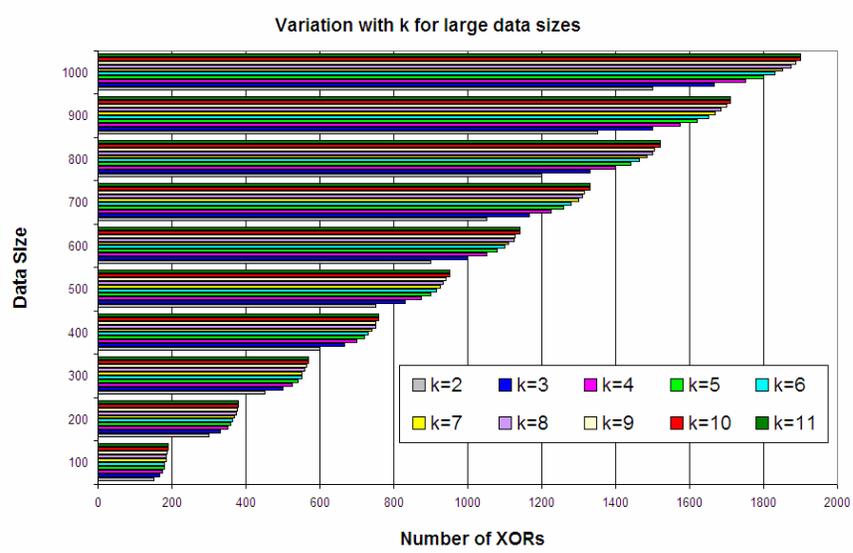

*Figure 2: Variation of data size with the number of XORs for different k*

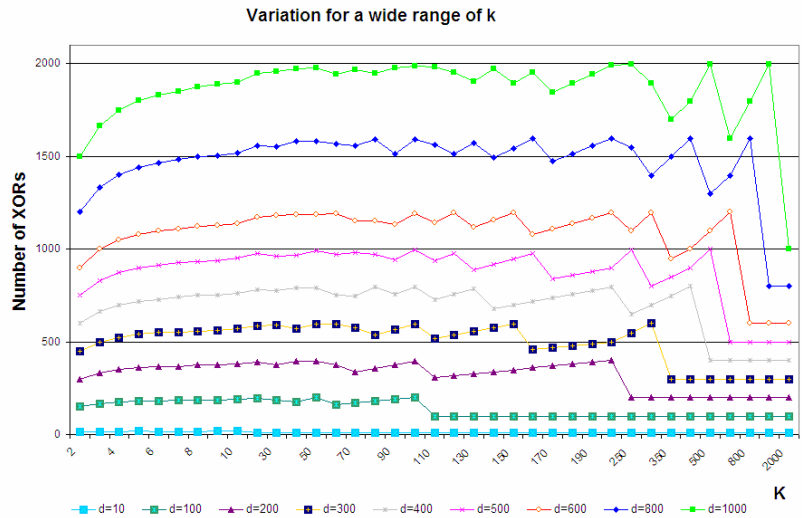

*Figure 3: Variation of number of XORs over k for large data sizes*

It all illustrates the picture of the number of XORs for a particular data size over different values of *k*. If the size of the key is large enough (that is, if *k* is large enough) to cover the entire data size in a single round, then the number of XORs required is drastically reduced. This is indicated by the sharp dip and then stability in the lines of different data sizes. Note that the scale in the x-axis is not uniform, and it is to bring out this behavior that the graph has been drawn on this scale. For values of *k* lesser than this "threshold" value, we see that the value of *k* = 2 is the point of minimum. (It is also to be noted here that *k* = 1 is not possible since in that case our algorithm would only be transferring random numbers and no data! Therefore, the least permissible integral value for *k* is 2).

The above figures concur that the optimum value for *k* to reduce the number of XOR operations in the algorithm is 2. In fact, for this value, the algorithm cannot be attacked using standard cryptanalytic attacks owing to the almost comparable sizes of the key (which is 2 * *PS* bits) and the random number used for the one-time pad implementation (which is *PS* bits). This is further explained in detail in the next sub-section. Also, the key to be exchanged between the sender-receiver pair is smallest for this permissible value of *k*. However, such a low value for *k* translates to a large value for the number of multiplications to be done as part of the algorithm as is shown in Figure 4.

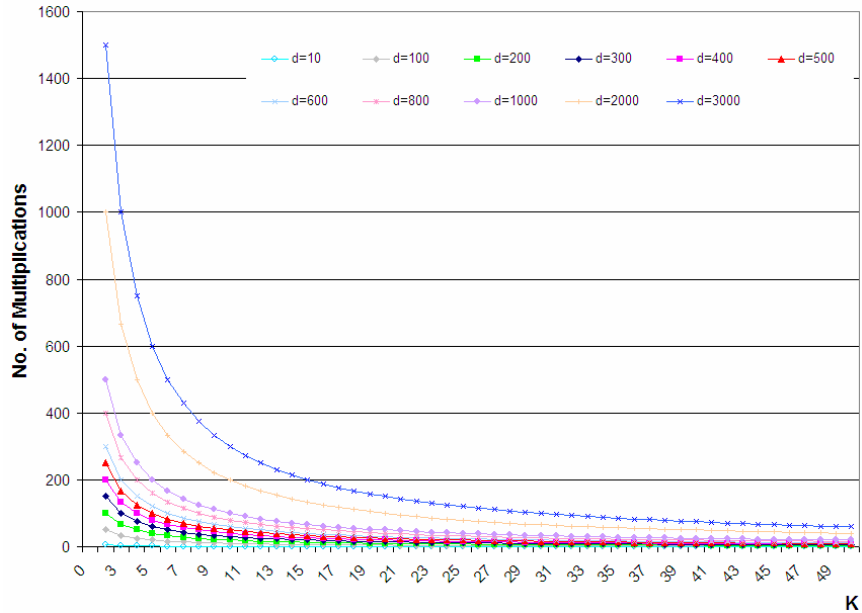

*Figure 4: Number of multiplications over a typical range of k for different data sizes*

Figure 4 depicts the behavior of the number of the multiplications, which is inversely proportional to the value of $k$. Hence, for small values of $k$, the number of multiplications is large and grows larger for larger data sizes. A point to be noted here is that these multiplications are field operations with inputs being *PS*-bit size blocks. Whereas in AES, all multiplicative-type operations are field operations with 8-bit sized input blocks, which is at least 2 orders of magnitude lesser than *PS*.

In conclusion on the optimal value of $k$, there is a tradeoff between the size of the key and the number of XOR operations on the one hand versus the number of multiplications on the other. Size of the input blocks for these operations is also an important factor. A suitable value for $k$ should be arrived at after taking into consideration the above statistics and behaviors.

Our algorithm lends itself to parallelism in implementation in software as well as dedicated hardware. The execution of the operation *transform (P,R)* should follow the processing of $k$ blocks. This sequential nature cannot be avoided. However, the processing of the blocks can be done in a parallel manner. In principle, the algorithm can be implemented efficiently using $k$ XOR gates, as shown in Figure 5. Here, the first three cycles of XORs are depicted with the respective inputs and outputs. If number of gates is also a constraint, for best performance, the value of $k$ should be decided accordingly. In fact, if the number of gates or the area in the hardware implementation is not a constraint, then the number of XOR gates could be doubled which would effectively double the performance of our algorithm. This is done by executing the operations otherwise performed in rounds I and II as illustrated in Figure 5 in the same round. (This can be done synchronously since $P$ is not changed until the second round and it is the common input to the first two rounds.) Note that, the units in the $k^{th}$ column of rounds II, IV and so on are actually *PS*-bit multipliers. Here, XORs have been placed purely to provide a picture of symmetry only and are not intended to misguide the reader.

**4.6 Security of the Jigsaw**

The security offered by the algorithm is the same as that provided by one-time pad - information-theoretic security. This is evident from the fact that the jigsaw is structured by an embedding of the data in the key using the XOR operation. However, unlike the one-time pad, keys used in our algorithm are not completely uncorrelated since the next key is formed as a function of the current key and a random value. Thus, even if some information about the key or the random number of a round or even about the nature of the data is revealed, the security offered by our approach fails while the theoretical one-time pad continues to offer the same level of security to the unexposed data.

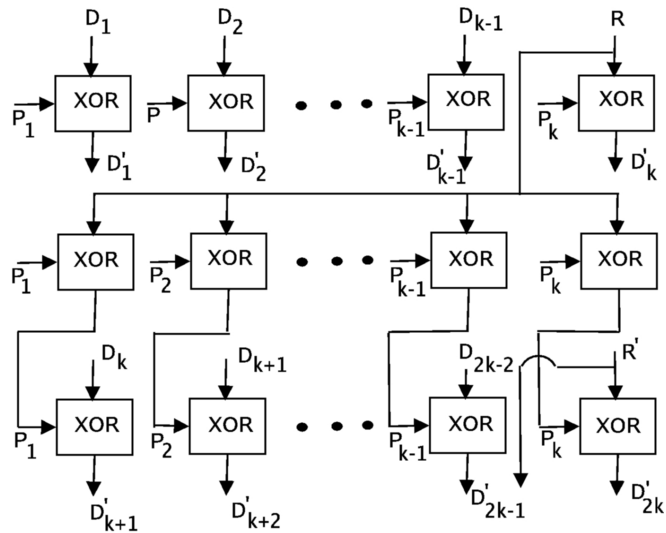

*Figure 5: Schematic for parallel implementation of the 'jigsaw' algorithm*

As an example, consider the case when the nature of the data that is being transferred is known to an adversary, which is very common in today's world either due to negligence, or due to the predictable nature of humans and machines or by other means. In such a case, here is an attack that could be carried out on the "jigsaw" algorithm. The first step is to guess the size of the key being used (that is, the value of $k$). Having guessed it, it is also important to figure out the position of the start of a new round in the stream of enciphered text captured by the adversary. Note that for every sender-receiver pair, the state information pertaining to the partial use of a key is stored at both the sender and receiver who therefore continue to use the same key for the next round of communication. Hence, for example, an adversary's intercepted data could actually start with the ending of a round and the beginning of the next and he might not have access to the end of the next round. This would render the following attack useless. Interception from the start of the communication between a sender-receiver pair is one way of eliminating this randomness.

Having guessed the key and demarcated the enciphered blocks of two consecutive rounds, the adversary could XOR the blocks of a round with the blocks of the immediately previous round in the corresponding positions. This would eliminate the secret key used in either of the rounds leaving the attacker with blocks that are of the form $D_i'$ *XOR* $D_j'$ *XOR R*, where $D_i'$, $D_j'$ are data blocks and *R* a random number, all of size *PS* bits. Now there are *k* such blocks available and therefore, by XORing two such blocks, the attacker can get rid of *R* and he is left with $D_i'$ *XOR* $D_j'$ *XOR* $D_k'$ *XOR* $D_l'$. If it is possible to guess the values of any of the data blocks at either this stage of XOR or the previous stage (knowing about the nature of the data), then the security offered by the algorithm is broken immediately.

Since the above eventuality is fairly common, it is necessary that we discuss a solution that makes such an attack at least computationally, if not theoretically, impossible. We suggest the use of some existing ideas in literature that do not harm the efficiency of our algorithm and at the same time offer protection against cryptanalytic attacks such as the ones illustrated above.

### 4.6.1 Supplementing "Jigsaw" with All-or-Nothing

The All-or-Nothing Transform (AONT) was first suggested by Rivest [Rivest, 1997]. We present below the definition of an AONT as given in [Rivest, 1997] with a slight modification. An AONT, say *f*, when applied on an input message sequence M to obtain the transformed message sequence M', has the following properties:

a. The transformation *f* is reversible; given M', one can obtain M.
b. Both the transformation *f* and its inverse $f^{-1}$ are efficiently computable (that is, in polynomial time).
c. It is infeasible to compute any function of any part of M if any part of M' is unknown.

However, the construction of the AONT in [Rivest, 1997], called the *package transform*, is only computationally secure. Stinson in [Stinson, 2001] had provided the definition of an AONT having the property of information-theoretic security. He had also provided a construction of a linear *(s, q)* - AONT that is unconditionally secure. This linear implementation of AONT has a non-randomizing property that did not rule out known- and chosen- plaintext attacks against it. So, Stinson had also suggested a method of randomizing the input to prevent such attacks. Thus, by randomizing input, the linear *(s, q)* - AONT prevents cryptanalytic attacks.

The approach in literature till now has been to use AONT as a preprocessing step before a standard encryption algorithm like DES or RSA. The reason is to slow down a brute force search of the encryption key. However, here we suggest the use of the AONT in the creation stage of our "jigsaw" puzzle where it is used to scramble the data to effectively remove any patterns present in the original data stream. In fact, the AONT is designed such that without the presence of all blocks of data in the proper order, it is impossible to gain any information about the original data. This in effect creates a "jigsaw" puzzle tougher than that available in the real world since, in the latter case, it is possible to inspect a piece of the puzzle and figure out where it should be placed.

### 4.6.2 Implementing AONT with Jigsaw Algorithm

In this paper, we use Stinson's implementation of the linear $(s, q)$ - AONT as given in Corollary 2.3 [Stinson, 2001]. A brief description of the algorithm is presented now. Suppose $q > 2$ is a prime power (of form $q = p^k$) and $s$ is a positive integer. Then there exists a linear $(s, q)$ - AONT whose construction is given as follows:

Given $x$, we can compute $y$ as:
1) For $1 \leq i \leq s - 1$, compute $y_i = x_i + x_s$.
2) Compute $x_s = x_1 + x_2 + ... + \lambda * x_s$.

And, given $y$, we can get back $x$ as:
1) Compute $y_s = \gamma * (x_1 + x_2 + ... + x_{s-1} - y_s)$.
2) For $1 \leq i \leq s - 1$, compute $x_i = y_i - x_s$.

where $x$, $y$ are messages of $s$ blocks each, $\lambda$ belongs to $\mathcal{T}_q$ such that $\lambda$ does not belong to $\{s - 1 \bmod p, s - 2 \bmod p\}$, and $\gamma = (s - 1 - \lambda)^{-1}$.

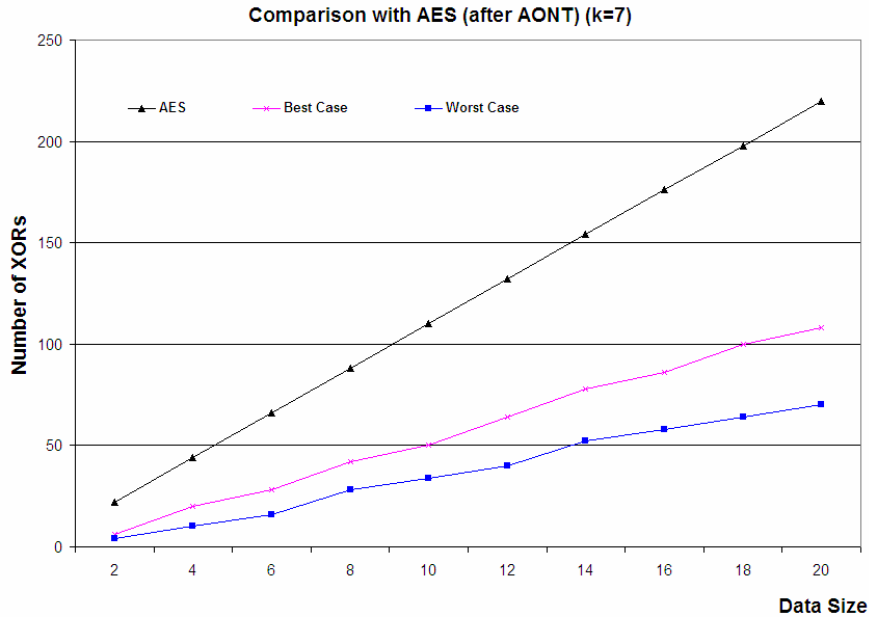

*Figure 6: Comparison of AES and jigsaw algorithm (including AONT) functioning in a) best case and b) worst case in the number of XORs*

In our algorithm, since *PS*, the size of a data packet, is typically a power of 2 (and therefore a prime power), we consider a *(N, $2^{PS}$)* - AONT for our algorithm. At any time, the data blocks created by the arbitrary tearing from the first step of the algorithm could be fed into this linear *(N, $2^{PS}$)* - AONT which produces an equal number of data blocks. Randomization of the input could be achieved by fixing the last block to be a random number and this information could be exchanged along with secret key. The modified algorithm would now require an additional *2 \* (N-1)* XORs and a *PS*-bit multiplication. Thus, in total, the new algorithm would require *3 \* N + floor(N/k) \* (k -1) - 2* additions and *floor(N/k) + 1* multiplications both on *PS*-bit sized inputs. Figure 6 depicts the number of XOR operations needed for the algorithm functioning in the best case and the worst case as was described earlier.

Compared with Figure 1, there has been a sizeable increase from the earlier algorithm (that is, without the AONT) while remaining well below the values for AES. As for the number of multiplications, the increase is only of constant in nature and the behavior of the line in figure 4 is not altered. Thus, as can be seen from the graphs and statistics, the addition of AONT in the creation step of the "jigsaw" algorithm does not make out algorithm inefficient computationally while providing a valuable addition to the security against cryptanalytic attacks.

To conclude, in the event of information leak regarding the data or the nature of it, our initial algorithm can be easily broken and a possible attack for the same was illustrated. However, with the use of the linear *(s, q)* - AONT, this security hole can be plugged very efficiently.

## 5  Future Research

Much research over the years has not yielded a practical implementation of the one-time pad. The 'jigsaw' algorithm is another drop in this ocean of research that seeks to realize this with our own set of disadvantages and limitations. In this respect, the main disadvantage with the algorithm is that an information leak due to human negligence has the potential to completely break the security of the algorithm. We believe that the paradigm of "jigsaw puzzle" as illustrated in this paper is novel and any research towards a sturdier implementation would benefit the community. In this paper, we have presented our implementation of the "jigsaw" paradigm. There could be other implementations that are more efficient than ours. More importantly, this paradigm could be better realized in a multi-dimensional way as opposed to the one-dimensional data stream model. To illustrate, consider the source data as a two-dimensional matrix of bits. Then, allowing arbitrary tearing along any direction would likely create the basis for a "2D-jigsaw puzzle". In such cases, alternative implementations for conveying puzzle information securely to the receiver(s) have to be evolved.

## 6  Conclusions

In this paper, we adopt a "jigsaw" approach to secure data transfer in networks. The data to be sent is broken into parts of arbitrary sizes forming the basis for creating a "jigsaw puzzle". Enough information is provided efficiently and securely to enable the receiver to solve the puzzle. The transfer of the parts is done securely without leaking any information to the adversary regarding the data. We have illustrated a method of implementing message authentication by private key without the exchange of any more information. The concept of the one-time pad is implemented in the course of the algorithm resulting in information-theoretic security of data transfer. The issue of key management is addressed by firstly the exchange of a large number a priori, and then subsequent modifications to the large number at regular intervals. These modifications are designed such that their outputs seem random to the adversary. Also, flexibility in the form of a means of control is provided in the algorithm to monitor and check the overhead resulting because of the data expansion due to the arbitrary splitting. Tradeoffs involved in the practical realization of the algorithm have been discussed and their relative impacts on the performance analyzed. We have also suggested a technique to make the algorithm secure against cryptanalytic attacks in the eventuality when the nature of the data is revealed. The inclusion of this has been proved to still be substantially more efficient than encryption algorithms. Moreover, our realization of the "jigsaw" paradigm has been designed to support a parallel implementation catering to future technological advancements.

**Acknowledgements**

Authors would like to thank the anonymous reviewers for their valuable comments which helped to improve the clarity of the paper.